\documentclass[12pt]{article} 

\usepackage{OL}
\usepackage{hyperref}
\usepackage{amsmath}
\usepackage{epsfig}
\graphicspath{{figs/}}

\begin{document}

\title{Air-guiding photonic bandgap fiber with improved triangular air-silica photonic crystal cladding}

\author{M. Yan, and P. Shum}

\address{yanmin@pmail.ntu.edu.sg \\ Network Technology Research Centre, Nanyang Technological University \\ 50 Nanyang Drive, Singapore, 637553}

\begin{center}
{\bf{Abstract}}
\end{center}
We introduce a small-core air-guiding photonic crystal fiber whose
cladding is made of improved air-silica photonic crystal with
non-circular air holes placed in triangular lattice. The fiber
achieves un-disturbed bandgap guidance over 350nm wavelength
range.

\section{Introduction}
\begin{figure}
\centering
\includegraphics[width=4cm]{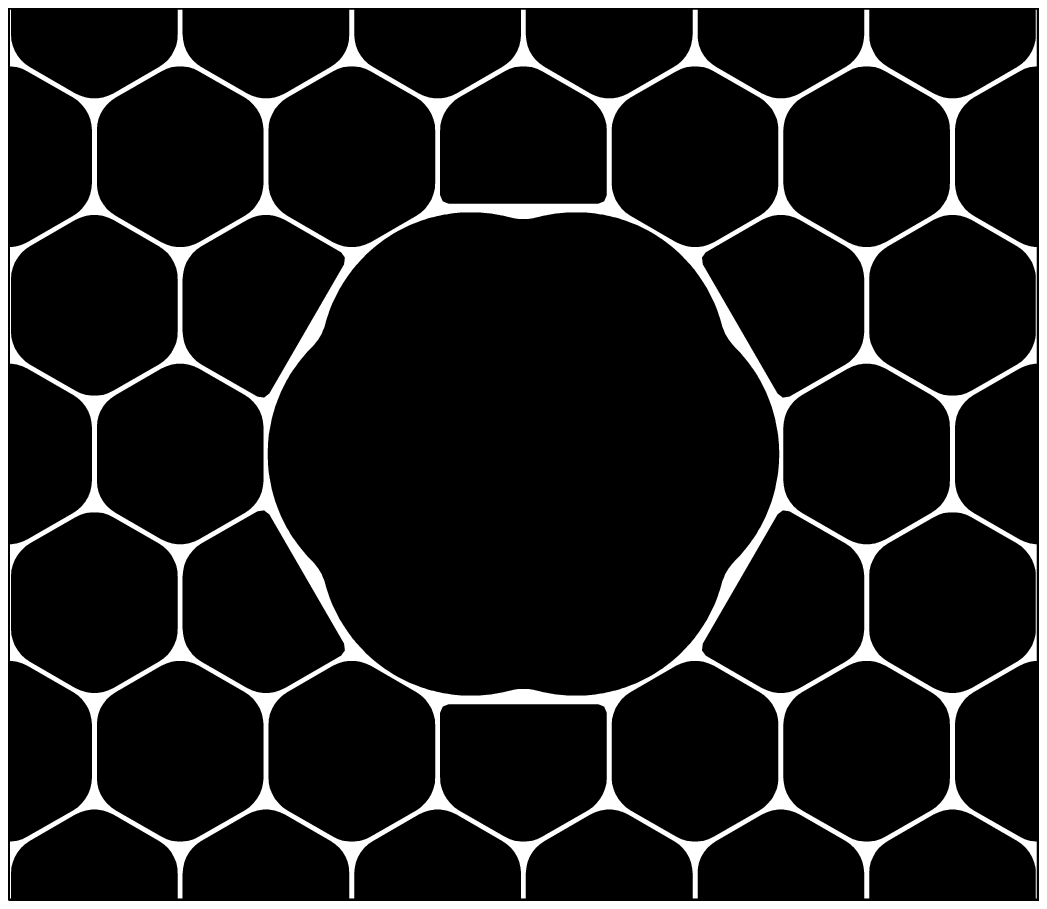}\ \ \ \ \ \ \
\includegraphics[width=4cm]{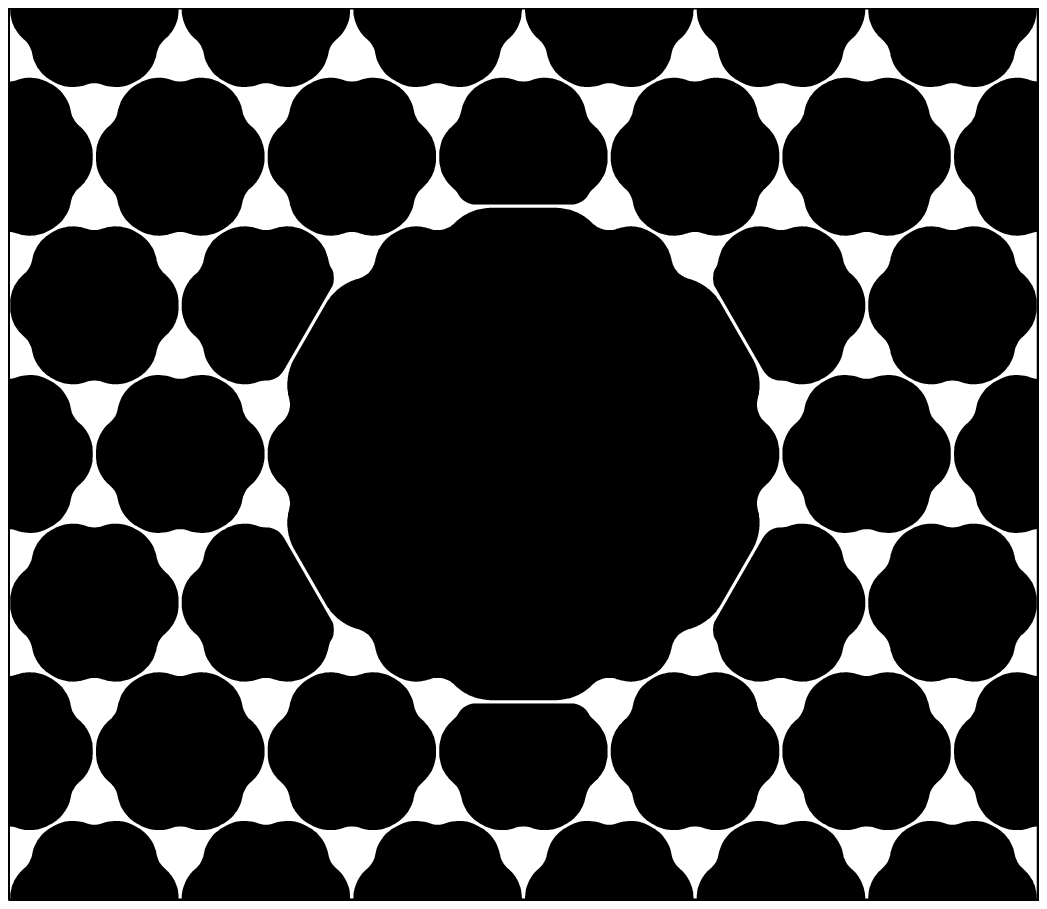}\ \ \ \ \ \ \
\includegraphics[width=4cm]{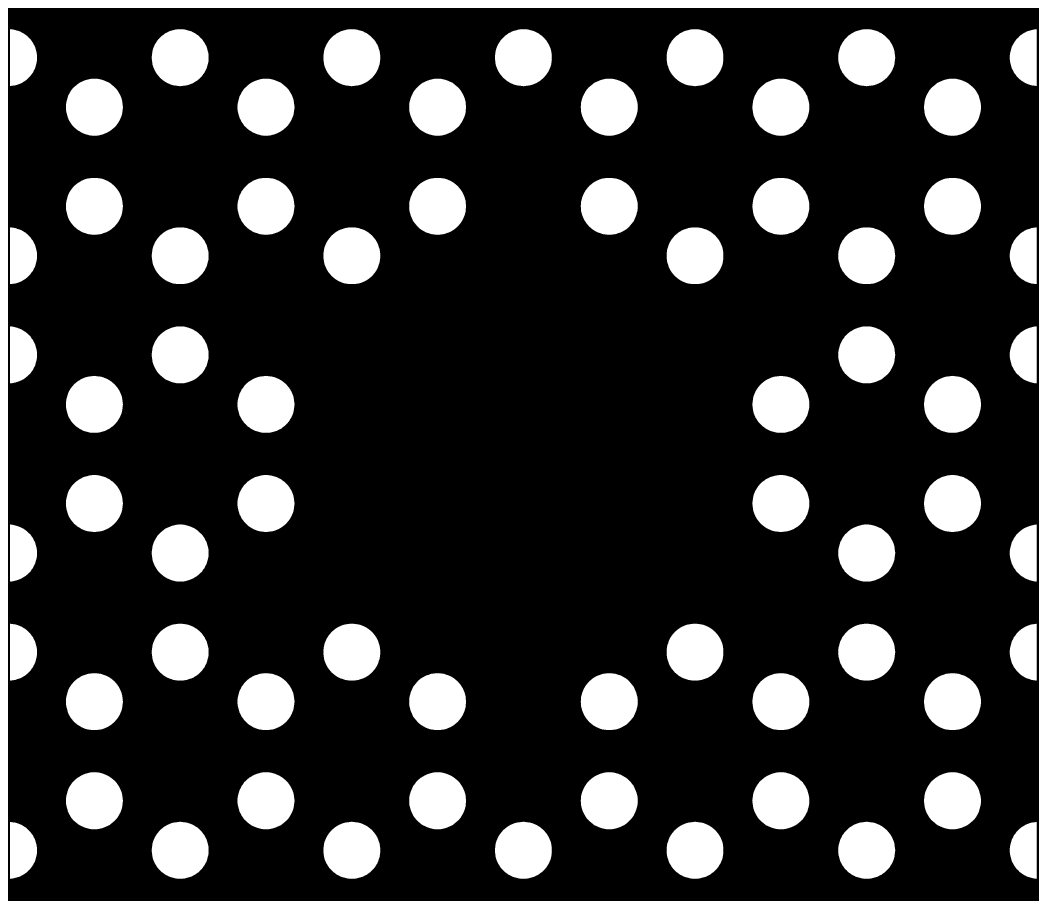}
\put(-345, -12){\textsf{(a)}} \put(-207, -12){\textsf{(b)}}
\put(-63, -12){\textsf{(c)}} \caption{(a) Air-guiding PCF reported
in Ref. \citeonline{Smith:TPCFCorning}. (b) Proposed air-guiding
PCF. (c) Ideal rods-in-air PCF. Black is for air.}
\label{fig:threePBGFs}
\end{figure}

Lightwave delivery in a hollow-core photonic crystal fiber (PCF)
is of significant importance for applications like laser beam
handling, nonlinear optics in gases, sensing, atom/particle
guiding, and even for low-loss optical communication etc
\cite{Russell:PCF,Smith:TPCFCorning}. Current air-guiding PCF
[Fig. \ref{fig:threePBGFs}(a)] explicitly uses air-silica photonic
crystal (PC) made from a bundle of thin silica tubes (triangular
placement) in its cladding \cite{Smith:TPCFCorning}. Such fiber
has two main disadvantages: surface-mode interference and
multimode operation. Though theoretically, there does exist some
design which eliminates the surface-mode problem
\cite{West:TPCFSurfaceMode}, such fiber hasn't been fabricated,
largely due to difficulties in preform preparation and maintaining
the core shape during drawing. In this paper, we introduce a
hollow-core PCF [Fig. \ref{fig:threePBGFs}(b)] whose cladding is
made of newly proposed air-silica PC \cite{Yan:TPCImproved}. The
improved PC allows designs with smaller core size and/or wider
transmission window, etc \cite{Yan:TPCImproved}. We will show that
the air-guided mode in our proposed fiber has an un-disturbed
transmission over 350nm wavelength range, owing to the fact that
the surface modes stay very close to the bandgap edge. Potential
single-mode operation with such type of hollow-core PCF is also
suggested.

\section{Design and Numerical Analysis}
\begin{figure}
\centering
\includegraphics[width=6cm]{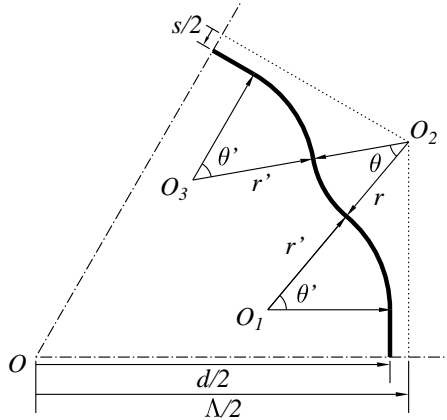}
\caption{Schematic diagram of a portion (1/6) of the improved PC
unit. Thick line is air-silica interface, with air to its left.}
\label{fig:PortionSketch}
\end{figure}

In Fig. \ref{fig:threePBGFs} we show three air-guiding PBG fibers.
In fact, both cladding PCs in Fig. \ref{fig:threePBGFs}(a) and (b)
are derived from the rods-in-air PC shown in Fig.
\ref{fig:threePBGFs}(c). Guiding mechanisms of three fibers are of
no difference --- all of them can guide light with a cladding
bandgap that exists between the PCs' 4th and 5th bands (computed
using the plane-wave method with primitive basis vectors); and we
find the mode profiles in low-order bands are equivalent for three
PCs. One portion of the cladding PC unit in Fig.
\ref{fig:threePBGFs}(b) is sketched in Fig.
\ref{fig:PortionSketch}. For the particular fiber shown in Fig.
\ref{fig:threePBGFs}(b), its cladding has $\Lambda=2.6\mu$m,
$d=0.98\Lambda$ ($s=0.02\Lambda$), $r=0.14\Lambda$, and
$\theta=40^\circ$. The core is formed by removing 12 silica
pillars. Though we can easily get rid of the surface-mode problem
theoretically by using a design rule suggested in
\cite{Digonnet:SurfaceModes}, we stick to a practical core shape
[Fig. \ref{fig:threePBGFs}(b)] to facilitate easy stacking and
pressurization. Extra silica veins surrounding air core are of
thickness $s$.

\begin{figure}
\centering
\includegraphics[width=9cm]{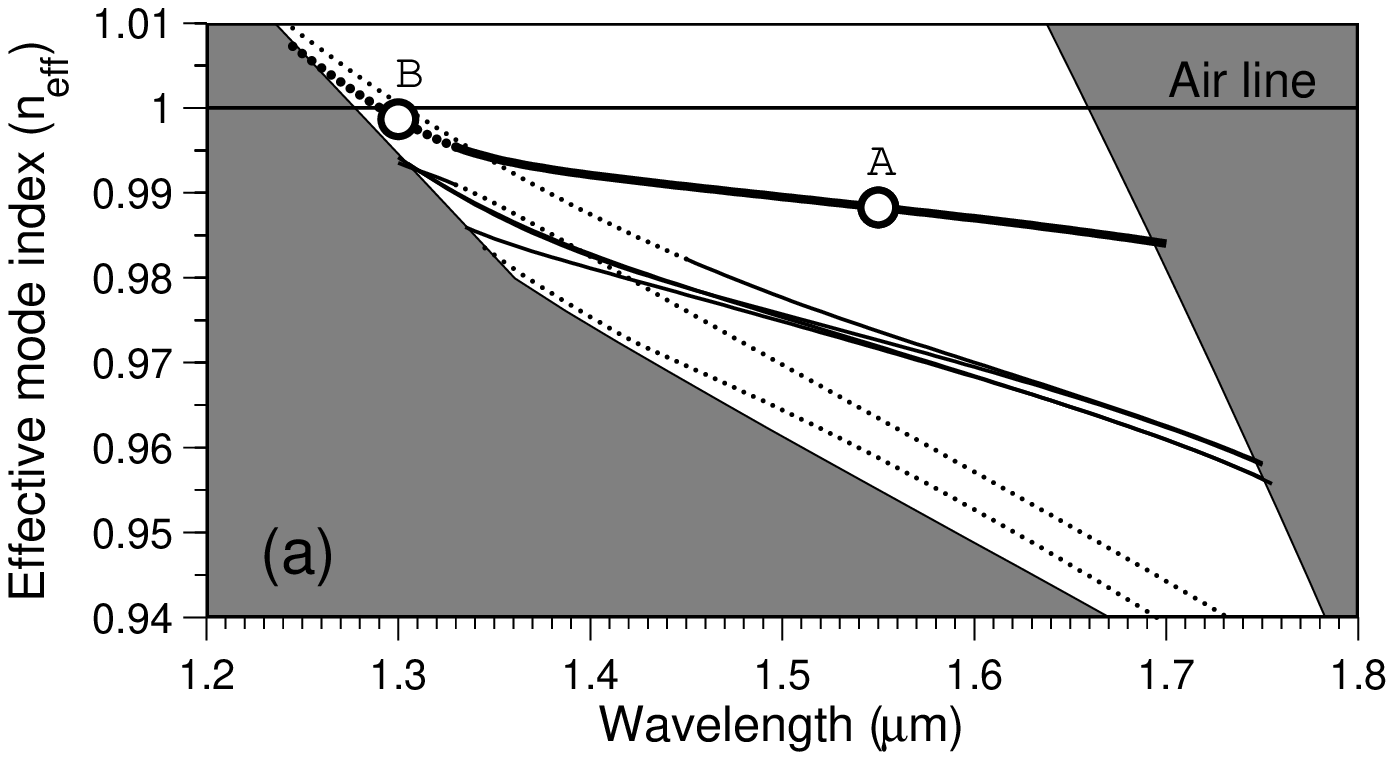}
\\ \
\\
\includegraphics[width=9cm]{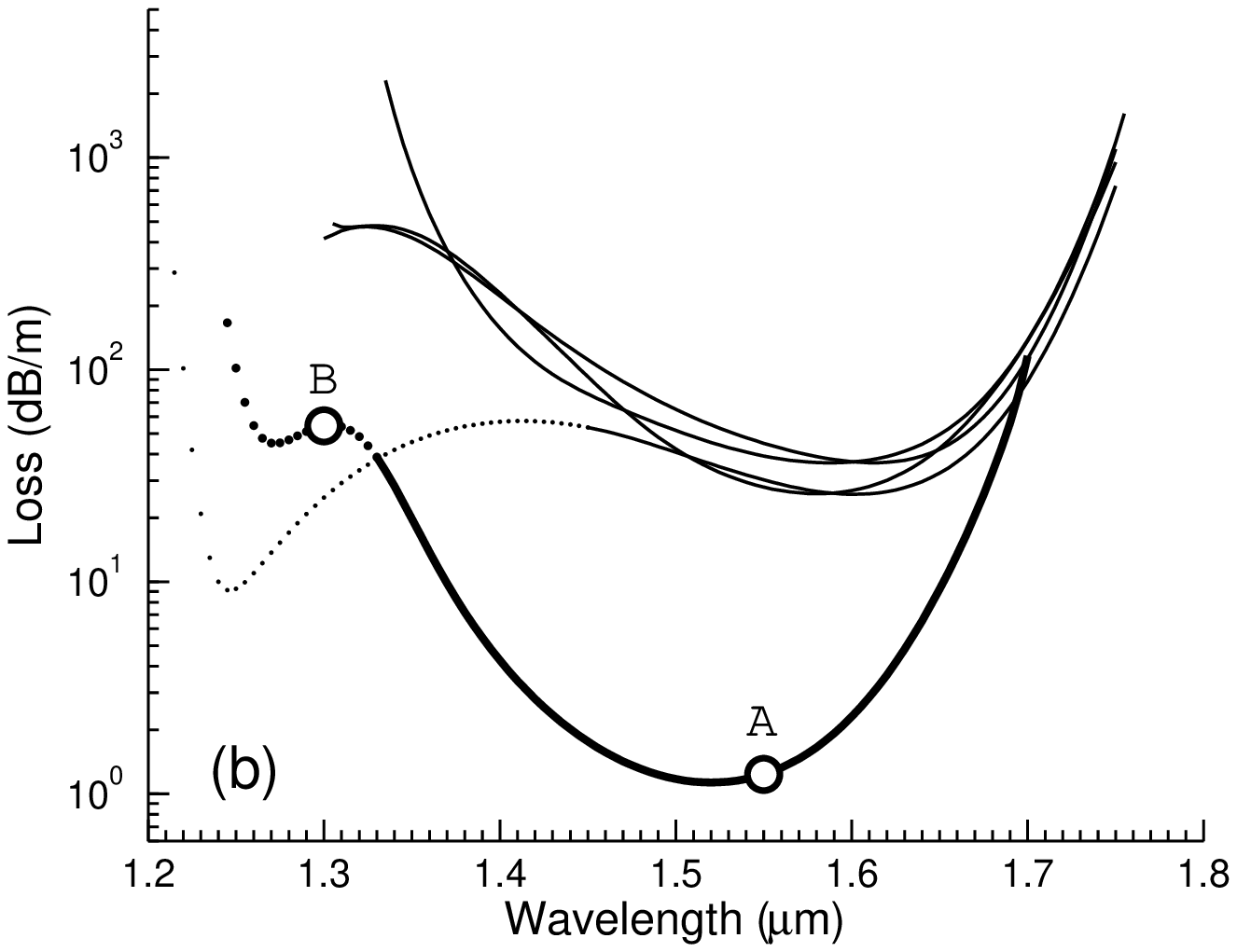}
\caption{Dispersion (a) and loss (b) curves of the defect modes.
Photonic bandgap region of the cladding PC is represented by the
white patch in (a).} \label{fig:DispersionLossTPCFAT}
\end{figure}

The photonic bandgap (PBG) region possessed by the cladding PC is
shown in Fig. \ref{fig:DispersionLossTPCFAT}(a) by the white
patch. It is noticed that the region is extending beyond air line
to $n_{\mbox{\small eff}}=0.922$ (not shown), which is
significantly smaller than the value achievable with PC in Fig.
\ref{fig:threePBGFs}(a) (0.968). This feature allows us to design
air-guiding fibers with smaller core and/or lower-loss air-guiding
PCFs. We then use a full-vector finite-difference mode solver
\cite{Guo:FDFD} to compute guided defect modes with four air-hole
rings in the cladding. Numerical resolution is at $dx=dy=0.12\mu$m
with $11\times{11}$ sub-grid index averaging. Perfectly matched
layers have 12-grid thickness. The two degenerate fundamental
air-guided modes ($\mbox{HE}_{11}$-like) are shown by the thick
solid curve in Fig. \ref{fig:DispersionLossTPCFAT}(a). It is found
the modes are un-disturbed in $1.35\sim{1.70}\mu$m wavelength
range. Their loss spectrum is shown by the thick curve in Fig.
\ref{fig:DispersionLossTPCFAT}(b). Minimum loss is about 1dB/m.

\begin{figure}
\centering
\includegraphics[height=5cm]{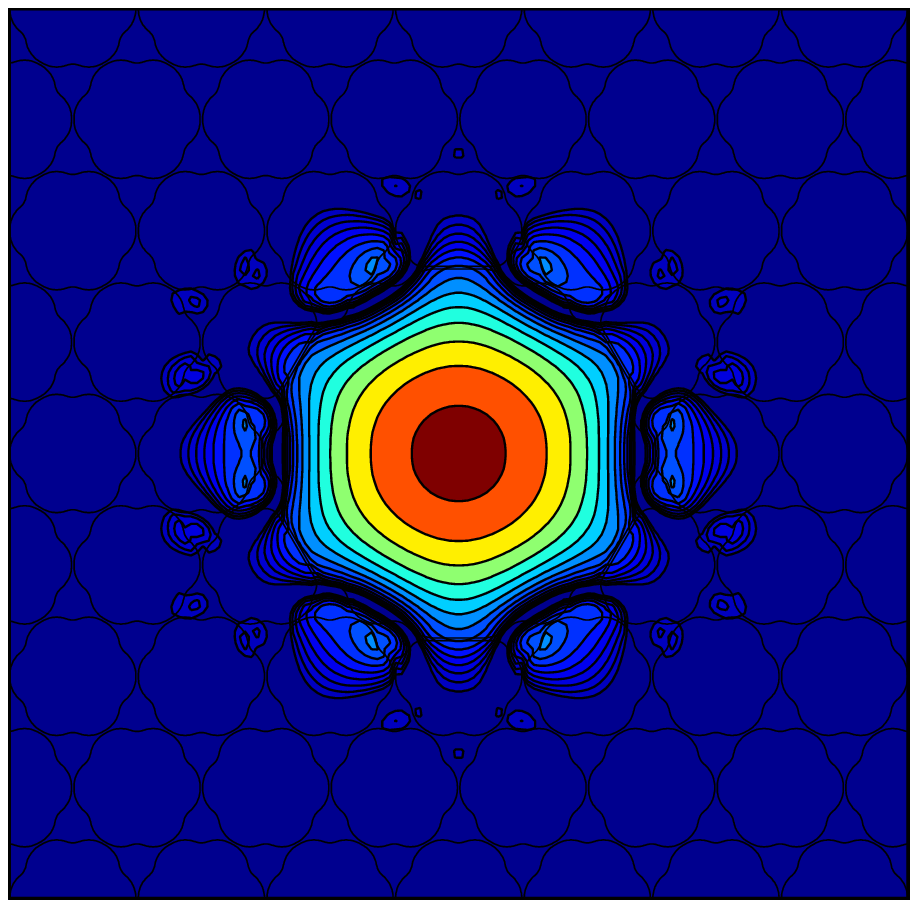} \ \ \ \ \ \ \ \
\ \ \ \
\includegraphics[height=5cm]{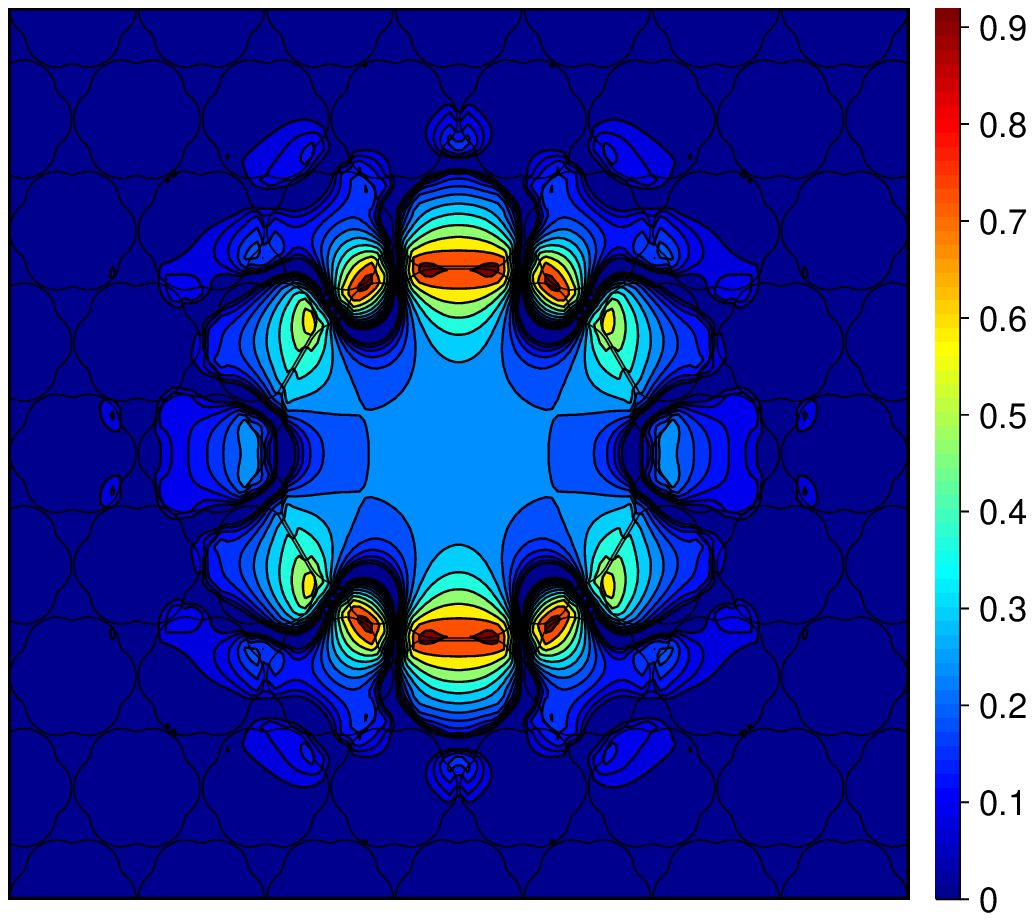}
\put(-290, -10){\textsf{(a)}} \put(-96, -10){\textsf{(b)}}
\caption{Air-guided mode at $\lambda=1.55\mu$m (a) and surface
mode at $\lambda=1.30\mu$m (b). Contour lines are in 1-dB
separation.} \label{fig:ModesTPCFAT}
\end{figure}

The $|\mbox{E}_x|$ field distributions of modes at point $A$
($\lambda=1.55\mu$m, air-guided mode) and $B$ ($\lambda=1.3\mu$m,
surface mode) are shown in Fig. \ref{fig:ModesTPCFAT}(a) and (b),
respectively. The core mode at $1.55\mu$m is very well confined,
and it has leakage loss of 1.2dB/m, which will decrease to
0.053dB/m and 0.003dB/m when the number of rings in cladding is
five and six, respectively. Though the six-ring loss value is
higher than that for the fiber given in Fig.
\ref{fig:threePBGFs}(a) \cite{Saitoh:TPCFLeakage}, it should be
reminded that our core size is significantly smaller (diameter
$\sim7.4\mu$m v.s. $\sim13.6\mu$m). Loss should be decreased if
additional pillars are removed in core region.

The proposed fiber is still multimode, largely because the gap
region extends quite far beyond air line. The dispersion curves of
second-order modes ($\mbox{TE}_{01}$-, two $\mbox{HE}_{21}$- and
$\mbox{TM}_{01}$-like modes) are shown in Fig.
\ref{fig:DispersionLossTPCFAT}(a) by four thin solid lines. Their
loss values (in dB/m), represented by thin lines in Fig.
\ref{fig:DispersionLossTPCFAT}(b), are about 30 times higher than
that of the fundamental modes. It should be noticed that, due to
the small core size, the dispersion curves of the fundamental and
second-order mode groups in Fig. \ref{fig:DispersionLossTPCFAT}(a)
stay further apart as compared to the fiber reported in
\cite{Smith:TPCFCorning}. This means the coupling between the two
mode groups is smaller. With careful excitation, we can achieve
single-mode operation for applications like laser beam delivery,
in which severe fiber bending can be purposely avoided.

Surface modes are denoted by dotted lines in Fig.
\ref{fig:DispersionLossTPCFAT}. An obvious advantage of our fiber
is that the surface modes stay very close to the bandgap edge. We
attribute this to the analogousness between silica pillars nearby
core and those in cladding. It is observed from Fig.
\ref{fig:ModesTPCFAT}(b) that a nodal line appears in each silica
pillar adjacent to core. Such surface modes are similar to
$5^{th}$-band bulk cladding mode. They are pulled into gap region
because the pillars nearby core have slightly more silica than
those in cladding PC. Hence their modal energy is lower (higher in
$n_{\mbox{\small eff}}$). By varying cladding PC parameters, we
should be able to further reduce the impact of these modes,
\emph{i.e.}, to push them closer to the gap region boundary.

\section{Conclusion}
We have proposed a low-loss air-guiding PCF design which has 350nm
un-disturbed PBG guiding wavelength range.

\section*{Acknowledgements}
M. Yan acknowledges Optical Fibre Technology Centre, University of
Sydney for providing computing facility. He is partially supported
by a scholarship provided by A*STAR, Singapore.

\bibliographystyle{osajnl}
\bibliography{PCFbib}

\end{document}